\documentclass[aps,twocolumn,superscriptaddress,showpacs]{revtex4}

\usepackage[dvips]{graphicx}  
\usepackage{color}  
\newcommand{\nc}{\newcommand}           
\nc{\vc}[1]     {\mbox{\boldmath $#1$}} 
\nc{\bra}       {\langle}               
\nc{\ket}       {\rangle}               
\nc{\bras}[1]   {\langle#1|}            
\nc{\kets}[1]   {|#1\rangle}            
\nc{\del}       {\partial}              
\nc{\red}[1]    {\textcolor{red}{#1}}  
\nc{\blue}[1]   {\textcolor{blue}{#1}}  
\nc{\green}[1]   {\textcolor{green}{#1}}  
\nc{\beq}     {\begin{eqnarray}}
\nc{\eeq}    {\end{eqnarray}}

\nc{\mydraft}	{\setlength{\topmargin}{-1.0cm}} 
\mydraft

\begin{document}

\title{Shell and alpha cluster structures in $^8$Be with tensor-optimized shell model}

\author{Takayuki Myo\footnote{myo@ge.oit.ac.jp}}
\affiliation{General Education, Faculty of Engineering, Osaka Institute of Technology, Osaka, Osaka 535-8585, Japan}
\affiliation{Research Center for Nuclear Physics (RCNP), Osaka University, Ibaraki, Osaka 567-0047, Japan}

\author{Atsushi Umeya\footnote{aumeya@nit.ac.jp}}
\affiliation{Human Science and Common Education, Faculty of Engineering, Nippon Institute of Technology, Saitama 345-8501, Japan}

\author{Kaori Horii\footnote{horii@rcnp.osaka-u.ac.jp}}
\affiliation{Research Center for Nuclear Physics (RCNP), Osaka University, Ibaraki, Osaka 567-0047, Japan}

\author{Hiroshi Toki\footnote{toki@rcnp.osaka-u.ac.jp}}
\affiliation{Research Center for Nuclear Physics (RCNP), Osaka University, Ibaraki, Osaka 567-0047, Japan}

\author{Kiyomi Ikeda\footnote{k-ikeda@postman.riken.go.jp}}
\affiliation{RIKEN Nishina Center, Wako, Saitama 351-0198, Japan}

\date{\today}

\begin{abstract}
We study the shell and $\alpha$ cluster structures in the ground and excited states of $^8$Be in terms of the tensor-optimized shell model (TOSM). 
In TOSM, the tensor correlation is optimized in the full space of 2p2h configurations involving high-momentum components.
The short-range correlation is treated with the unitary correlation operator method (UCOM). 
We use the effective interaction based on the bare nucleon-nucleon interaction AV8$^\prime$.
The $^8$Be states consist of two groups of ground band states and highly excited states with the isospin $T$=0 and $T$=1.
It is found that the tensor contributions of the ground band states are stronger than the highly excited states and that the kinetic energies and the central contributions of the ground band states
are almost twice the $^4$He values. These features suggest two-$\alpha$ clustering for the ground band states in $^8$Be.
We also estimate the correlation energy of the $\alpha$ clustering using the $\alpha$ cluster model.
In the highly excited states, the calculated spectrum in TOSM reproduces the experimental level order and the relative energies of each level. 
This agreement suggests that those states can be interpreted as shell-like states.
The level order is found to be sensitive to the presence of the tensor force in comparison with the results using the Minnesota effective interaction without the tensor force.
It is also found that the tensor contributions in the $T$=0 states are stronger than the $T$=1 states, which is consistent with the state dependence of the tensor force.
\end{abstract}


\maketitle

\section{Introduction}

One of the important issues in nuclear physics is to understand the nuclear structure in connection with the properties of the nucleon-nucleon ($NN$) interaction~\cite{akaishi86,kamada01,arai11}. 
The $NN$ interaction has a strong tensor force at long and intermediate distances caused by the pion exchange and a strong central repulsion at short distances caused by the quark dynamics.0
Recently, it became possible to calculate nuclei up to a mass around $A\sim 12$ using an $NN$ interaction with the Green's function Monte Carlo method~(GFMC) \cite{pieper01, pudliner97}. 
At present, this method requires a great deal of computational time to be applied to heavier nuclei. 
It is desirable to develop a new method to calculate nuclear structure with large nucleon numbers by taking care of the characteristic features of the $NN$ interaction.

The presence of the tensor force in the $NN$ interaction induces $d$-wave components in a nucleus, in particular, for proton-neutron ($pn$) pairs as a deuteron.  
The $d$-wave component causes the deuteron to be bound via the large $sd$ coupling of the tensor force. 
This $d$-wave component is found to be spatially compact as compared with the $s$-wave one due to large momentum characters brought by the tensor force \cite{ikeda10}. 
This effect originates from the pseudo-scalar nature of the one-pion exchange.  
It has been reported experimentally that a large fraction of $pn$ pairs is observed as compared with $pp$ or $nn$ pairs in the knockout of nucleons with momenta larger than the Fermi momentum for light nuclei \cite{subedi08, simpson11}.  
This enhancement of $pn$ pairs is hard to reproduce theoretically in a simple shell model \cite{simpson11}, except for the rigorous method such as the GFMC one \cite{schiavilla07}, 
which treats the tensor force explicitly. 
It is important to make efforts to study high-momentum components caused by the tensor force in finite nuclei {\cite{tanihata10}.
A recent experiment has observed the signature of the tensor correlation from the high-momentum component of the nucleon in the nucleus using the ($p$,$d$) reaction~\cite{ong13}.

The tensor force also contributes to making the $\alpha$ particle strongly bound by about 7 MeV per nucleon. It is known that the tensor contribution in the $\alpha$ particle is generally large~\cite{akaishi86,kamada01}, although its amount depends on the choice of the nucleon-nucleon interactions.
In light nuclei, it is often observed that the $\alpha$ particles are strongly developed as a cluster, e.g., in $^8$Be and the Hoyle state of $^{12}$C \cite{ikeda68,horiuchi12}.
For $^8$Be, the ground, $2^+_1$, and $4^+_1$ states are regarded as states consisting of two weakly interacting $\alpha$ particles. 
On the other hand the excited states above $4^+_1$ can be considered to have shell-like structures, because the $\alpha$ decay is not always favored.
For $^{12}$C, the ground state is rather a shell-like state and some of the excited states including the Hoyle state are recognized as a triple $\alpha$ cluster state,
part of which has an $\alpha$ condensate state nature \cite{tohsaki01,funaki02}.
It is interesting to discuss the coexistence of the $\alpha$ cluster state and the shell-like state in light nuclei.
In particular, we focus on the role of the tensor force in this phenomenon.

There are two important developments for performing nuclear structure calculations using the bare $NN$ interaction. 
One development is to find out that the strong tensor force is of intermediate range and 
we are able to express the tensor correlation in a reasonable shell model space~\cite{myo07,myo09,myo11,myo12}.  
We name this method as tensor-optimized shell model (TOSM).  
The other is the unitary correlation operator method (UCOM) to treat the short-range correlation caused by the short-range repulsion~\cite{feldmeier98, neff03, roth10}. 
We shall combine these two methods, TOSM and UCOM, to describe nuclei using a bare $NN$ interaction.  
In the TOSM part, the wave function is constructed in terms of the shell model basis states with full optimization of two particle-two hole (2p2h) states. 
There is no truncation of the particle states in TOSM, where spatial shrinkage of the particle states is essential to obtain convergence of the tensor contribution 
involving high-momentum components~\cite{toki02,sugimoto04,ogawa06}.
This treatment of the bare tensor force in TOSM corresponds to the one-pair approximation correlated by the tensor force~\cite{togashi07,ogawa11,horii12}.  

So far, we have obtained successful TOSM results for the investigation of the tensor correlations in He and Li isotopes.
In $^4$He, we showed the selectivity of the $(p_{1/2})^2(s_{1/2})^{-2}$ configuration of the $pn$ pair in the 2p2h space, which is induced by the tensor force.
This correlation was recognized as a deuteron-like state \cite{myo09,horii12}.  
The specific 2p2h excitations play a decisive role in reproducing the spectra of neutron-rich He and Li isotopes \cite{myo11,myo05,myo06,myo12}.
In He isotopes, the $p_{3/2}$ occupation of extra neutrons increases the tensor correlations of nuclei from $^5$He to $^8$He, 
while the $p_{1/2}$ occupation of extra neutrons decreases the tensor correlation of those nuclei
due to the Pauli blocking between the specific 2p2h excitations by the tensor force in $^4$He and the motions of the extra $p_{1/2}$ neutrons. 
The configuration dependence of the tensor correlation produces the right amount of splitting energy between the $p_{3/2}$- and $p_{1/2}$-dominant states in the He and Li isotopes.
We have also confirmed the importance of this blocking effect on the halo formation in $^{11}$Li \cite{myo07_11}.

In this paper, we apply TOSM to the $4N$ nuclei, $^{4}$He and $^8$Be, and see how the TOSM describes $^8$Be, 
which has different structures of the shell model and $\alpha$ clustering types appearing in the different excitation energy regions. 
In particular, we pay attention to the roles of the tensor force in determining those structures of $^8$Be.
Experimentally, the $^8$Be energy spectrum shows two groups.
One has the three ground band states of $0^+_1$, $2^+_1$ and $4^+_1$ as the rotational band. 
These states decay into two $\alpha$ particles and are understood as the two $\alpha$ cluster states in the intrinsic structure \cite{ikeda68,pieper04}.
The other group is the highly excited states starting from the $2^+_2$ state at $E_x=16.6$ MeV. 
Above this state, many spin states have been observed and the decay processes of these states are not only $\alpha$ emission,
but also the emissions of a proton, a neutron and $\gamma$. In addition, the decay widths of these states are almost less than 1 MeV, much smaller than the values of the $2^+_1$ and $4^+_1$ ground band states.
In the highly excited states, the $T$=1 states are also observed and degenerated with the $T$=0 states. They are considered as isobaric analog states of $^8$Li and $^8$B.
These experimental facts about $^8$Be indicate that the internal structures of the highly excited states are quite different from the three ground band states. 
In this sense, the $^8$Be nucleus possesses different features in the ground band and highly excited states.
It is theoretically known that a single $\alpha$ particle contains the strong tensor correlation and thus it would be interesting to see how the tensor force affects the variety of structures of $^8$Be.

In this study, we investigate the positive parity states of $^8$Be using TOSM and clarify the different structures of $^8$Be mentioned above from the viewpoint of the tensor force.
For this purpose, we use the effective interaction based on the bare $NN$ interaction AV8$^\prime$ , which is defined to simulate almost exactly the few-body results of $^4$He using the TOSM wave function,
retaining the characteristics of the bare $NN$ interaction as much as possible.
We also investigate how the non-central tensor and $LS$ forces determine the energy spectrum of $^8$Be.
For the $\alpha$ clustering, we discuss how well TOSM describes the two $\alpha$ components of $^8$Be and also estimate the explicit correlation energy of the $\alpha$ clustering using the $\alpha$ cluster model.

In Sect.\,\ref{sec:model}, we explain the methods of the TOSM and UCOM approach. 
In Sect.\,\ref{sec:result}, we show the results of $^4$He and $^8$Be and discuss their characteristics as members of the $4N$ nuclei in relation to the tensor force.
A summary is given in Sect.\,\ref{sec:summary}.

\section{Theoretical framework}\label{sec:model}

\subsection{Tensor-optimized shell model (TOSM)}

We explain the tensor-optimized shell model (TOSM) for open shell nuclei. We begin with writing a many-body Hamiltonian for an $A$ body system as
\begin{eqnarray}
    H
&=& \sum_i^{A} T_i - T_{\rm c.m.} + \sum_{i<j}^{A} V_{ij} , 
    \label{eq:Ham}
    \\
    V_{ij}
&=& v_{ij}^C + v_{ij}^{T} + v_{ij}^{LS} + v_{ij}^{Clmb} .
\end{eqnarray}
Here, $T_i$ is the kinetic energy of each nucleon with $T_{\rm c.m.}$ being the center-of-mass kinetic energy.  
We take a bare interaction $V_{ij}$ such as AV8$^\prime$ \cite{pudliner97} consisting of central $v^C_{ij}$, tensor $v^T_{ij}$, and spin-orbit $v^{LS}_{ij}$ terms, and $v_{ij}^{Clmb}$ the Coulomb term.  
We obtain a many-body wave function $\Psi$ by solving the Schr\"odinger equation $H \Psi=E \Psi$.  
In our previous work on $^{4}$He and a few-body systems, we found that the tensor force can be described by taking 2p2h excitations with large momentum components in the shell-model framework \cite{myo09,horii12}.

First, we prepare a standard shell-model state with $A$ nucleons in order to introduce the TOSM for open shell nuclei. The standard shell-model state $\Psi_{S}$ for $p$ shell nuclei is defined as
\beq
\Psi_{S}=\sum_{k_{S}} A_{k_{S}}|(0s)^{4}(0p)^{A-4};k_{S}\ket~.
\eeq
Here, the $s$ shell is closed, the $p$ shell is the valence shell, and the index $k_S$ is used to distinguish various shell-model configurations.
The concept of the TOSM is that the tensor force works strongly for two nucleons in the standard shell-model states and excites two nucleons to various two-particle states with high-momentum components.  
Hence, we limit configurations up to the two particle-two hole excitations from the standard shell-model states.

We take all the necessary high-momentum components brought by the strong tensor force. These components are included by exciting two nucleons from the $sp$ shells to higher shells 
as the 2p2h states from the standard shell-model state $\Psi_{S}$.  Hence, we have
\beq
\kets{{\rm 2p2h};k_{2}}=|(0s)^{n_{s}}(0p)^{n_{p}}({\rm higher})^{2};k_{2}\ket
\eeq
with the constraints $n_{s}+n_{p}=A-2$ and $2 \le n_{s} \le 4$.  We introduce the index $k_{2}$ to distinguish various 2p2h states, which amount to a large number of configurations.
Here, ``higher'' indicates higher shells outside of the $sp$ shells, which are treated as particle states in TOSM.
We use the notation of 2p2h states to specify that two particles are in the higher shells outside of the $sp$ shells. 

In addition to the 2p2h states with high-momentum component,
we extend the standard shell-model states by allowing two particles in the $s$ shell to excite into the $p$ shell in order to treat part of the tensor correlation in the $sp$ shells.  
We define these extended shell-model states as $\kets{{\rm 0p0h};k_{0}}$ states, since no particle is excited into the orbits of the particle states outside of the $sp$ shells.   
We express the extended shell-model states as
\beq
\kets{{\rm 0p0h};k_0}=|(0s)^{n_{s}}(0p)^{n_{p}};k_{0}\ket
\eeq
with the constraints $n_{s}+n_{p}=A$ and $2 \le n_{s} \le 4$.  The index $k_{0}$ distinguishes various $sp$ shell configurations. 
We can take into account part of the tensor correlation in the 0p0h states.

We also allow the 1p1h excitations for shell model consistency.
\beq
\kets{{\rm 1p1h};k_{1}}=|(0s)^{n_{s}}(0p)^{n_{p}}({\rm higher})^{1};k_{1}\ket
\eeq
with the constraint $n_{s}+n_{p}=A-1$ and $2 \le n_{s} \le 4$.  These 1p1h states can include high-momentum components and also improve the 0p0h wave functions in the radial components.

Finally, we define the TOSM wave function $\Psi$ for open shell nuclei as 
\begin{eqnarray}
\Psi&=& \sum_{k_0} A_{k_0} \kets{{\rm 0p0h};k_0} + \sum_{k_1} A_{k_1} \kets{{\rm 1p1h};k_1}
\nonumber\\
&+& \sum_{k_2} A_{k_2} \kets{{\rm 2p2h};k_2}.
      \label{eq:config}
\end{eqnarray}
Here, all the amplitudes $\{A_{k_0},A_{k_1},A_{k_2}\}$ are variational coefficients to be fixed by the energy minimization.

We explain the details of the radial wave functions. The 0p0h states are shell model states and expressed in terms of harmonic oscillator wave functions.  
Hence, the $0s$ and $0p$ shell-model wave functions are used to express the radial wave functions, whose length parameters are taken independently as variational parameters.
The 1p1h and 2p2h states involve particle states with high-momentum components. The hole states in those states correspond to the shell-model states in the $sp$ shells. 
The particle wave functions have to contain high-momentum components to express the specific effect of the tensor force with all the possible angular momenta until the total energy converges.

For particle states, we employ the Gaussian wave functions to express single-particle states in higher shells in order to describe high-momentum properties due to the tensor force~\cite{hiyama03, aoyama06}.  We prefer the Gaussian wave functions over the shell-model states in order to effectively include the necessary high-momentum components \cite{myo09}.
When we superpose a sufficient number of Gaussian wave functions with various length parameters, the radial components of the particle states can be fully expressed.
Gaussian basis states should be orthogonalized to the hole states and among themselves.
This condition is imposed by using the Gram-Schmidt orthonormalization. 
In order to use the non-orthogonal Gaussian basis functions in the shell model framework, we construct the following orthonormalized single-particle basis function $\psi^n_{\alpha}$ using a linear combination of Gaussian bases $\{\phi_\alpha\}$ with length parameter $b_{\alpha,\nu}$:
\begin{eqnarray}
        \psi^n_{\alpha}(\vc{r})
&=&     \sum_{\nu=1}^{N_\alpha} d^n_{\alpha,\nu}\ \phi_{\alpha}(\vc{r},b_{\alpha,\nu}),
        \label{eq:Gauss1}
        \\
        \bra \psi^n_{\alpha} | \psi^{n'}_{\alpha'}\ket
&=&     \delta_{n,n'}\ \delta_{\alpha,\alpha'},
        \label{eq:Gauss3}
	\\
{\rm for}~~n~&=&~1,\cdots,N_\alpha,       
        \nonumber
\end{eqnarray}
where $N_\alpha$ is a number of basis functions for the orbit $\alpha$, and $\nu$ is an index to distinguish the bases with a Gaussian length of $b_{\alpha,\nu}$.
The explicit form of the Gaussian basis function is written as
\begin{eqnarray}
        \phi_{\alpha}(\vc{r},b_{\alpha,\nu})
&=&     N_l(b_{\alpha,\nu}) r^l e^{-(r/b_{\alpha,\nu})^2/2} [Y_{l}(\hat{\vc{r}}),\chi^\sigma_{1/2}]_j ,
        \label{eq:Gauss2}
        \\
        N_l(b_{\alpha,\nu})
&=&     \left[  \frac{2\ b_{\alpha,\nu}^{-(2l+3)} }{ \Gamma(l+3/2)}\right]^{\frac12},
\end{eqnarray}
where $l$ and $j$ are the orbital and total angular momenta of the basis states, respectively. 
The weight coefficients $\{d^n_{\alpha,\nu}\}$ are determined to satisfy the overlap condition in Eq.~(\ref{eq:Gauss3}). 
This is done by using the Gram-Schmidt orthonormalization. 
Following this procedure, we obtain the new single-particle basis states $\{\psi^n_{\alpha}\}$ in Eq.~(\ref{eq:Gauss1}) used in TOSM. 
The particle states in 1p1h and 2p2h states are prepared to specify the basis wave functions, whose amplitudes are determined by the variational principle.  

We note that we can use another method to obtain $\{\psi^n_{\alpha}\}$ by solving the eigenvalue problem of the norm matrix of the Gaussian basis set in Eq.~(\ref{eq:Gauss2}) with the dimension $N_\alpha$.
This method gives different coefficients $\{d^n_{\alpha,\nu}\}$ from those of the Gram-Schmidt method. However, these two methods of making the orthonormalized single-particle basis states provide equivalent variational solutions for the total TOSM wave function $\Psi$ in Eq.~(\ref{eq:config}), because we start from the same Gaussian basis functions in Eq.~(\ref{eq:Gauss2}) with the same number $N_\alpha$ of Gaussian basis states.
Therefore, the TOSM solution does not depend on the construction methods of the orthonormal basis states using the Gaussian basis functions.

We construct Gaussian basis functions of particle states to be orthogonal to the occupied states.
For the $1s$ orbit in the particle states, we prepare an extended $1s$ basis function \cite{myo07,myo09}, which is orthogonal to the $0s$ state and possesses a length parameter $b_{1s,\nu}$ that can differ from $b_{0s}$ of the $0s$ state.
In the extended $1s$ basis functions, the polynomial part is changed from the usual $1s$ basis states to satisfy the conditions of the normalization and the orthogonality to the $0s$ state \cite{myo07}. 
For the $1p$ orbits, we take the same method as used for the $1s$ case. 
In the numerical calculation, we prepare at most 10 Gaussian basis functions with various range parameters to get a convergence of the energy and Hamiltonian components.
A typical value of the range parameters of the Gaussian base is from 0.3 fm to 6 fm.  

We note here that when we give the probabilities and occupation numbers of each orbit in various states in the numerical sections, 
those numbers are given by the summation of all the orthogonal orbits with the same spin having different radial behaviors, due to the fact these wave functions are constructed by orthonormalization.
For hole states, we do not have to sum up their numbers because the hole states are described by the single harmonic oscillator basis states.

We take care of the center-of-mass excitations by using the well tested method of introducing a Hamiltonian of center-of-mass motion in the many-body Hamiltonian known as the Lawson method~\cite{lawson}. 
In the present study, we take the value of $\hbar \omega$ for the center-of-mass motion as the averaged one used in the $0s$ and $0p$ orbits in the 0p0h states with the weight of the occupation numbers in each orbit \cite{myo11}.  
Adding this center-of-mass Hamiltonian as the Lagrange multiplier to the original Hamiltonian in Eq.~(\ref{eq:Ham}), we can effectively project out only the lowest HO state for the center-of-mass motion.  

The variation of the energy expectation value with respect to the total wave function $\Psi$ in Eq.~(\ref{eq:config}) is given by
\begin{eqnarray}
\delta\frac{\bra\Psi|H|\Psi\ket}{\bra\Psi|\Psi\ket}&=&0\ ,
\end{eqnarray}
which leads to the following equations:
\begin{eqnarray}
    \frac{\del \bra\Psi| H - E |\Psi \ket} {\del b_{\alpha,\nu}}
&=& 0\ ,\quad
   \label{eq:vari1}
    \\
    \frac{\del \bra\Psi| H - E |\Psi \ket} {\del A_{k_i}}
&=&  0\qquad \mbox{for}~i=0,1,2 .
   \label{eq:vari2}
\end{eqnarray}
The total energy is represented by $E$.
We solve two kinds of the variational equations in Eqs. (\ref{eq:vari1}) and (\ref{eq:vari2}) in the following steps.  
First, fixing the length parameters $b_{\alpha,\nu}$ and the partial waves of the basis states up to $L_{\rm max}$, 
we solve the linear equation for $\{A_{k_i}\}$ as an eigenvalue problem for $H$. 
We thereby obtain the eigenvalue $E$, which is a function of $\{b_{\alpha,\nu}\}$ and $L_{\rm max}$. 
Next, we try to adopt various sets of the length parameters $\{b_{\alpha,\nu}\}$, e.g., by changing the range (minimum and maximum values) of the length parameters in order to find a better solution that minimizes the total energy $E$ \cite{myo07}. 
Similarly we also increase $L_{\rm max}$ to see the convergence of the solutions.
In TOSM, we can describe the spatial shrinkage of particle states with an appropriate radial form in the individual configuration, 
which is important to describe the tensor correlation \cite{myo07}, as seen in the deuteron.
In this paper, we take $L_{\rm max}$ as 12 for the convergence of the numerical results.

\subsection{Unitary correlation operator method (UCOM)}

We briefly explain UCOM for the short-range central correlation \cite{feldmeier98,neff03,roth10}, 
in which the following unitary operator $C$ is introduced
\begin{eqnarray}
C     &=&\exp(-i\sum_{i<j} g_{ij})~.
\label{eq:ucom}
\end{eqnarray}
We express the correlated wave function $\Phi$ in terms of a simple wave function $\Psi$ as $\Phi=C\Psi$. 
The transformed Schr\"odinger equation becomes $\hat H \Psi=E\Psi$ where the transformed Hamiltonian is given as $\hat H=C^\dagger H C$.   
The operator $C$ is in principle a many-body operator. 
In the case of the short-range correlation, we are able to truncate the modified operators at the level of two-body operators~\cite{feldmeier98}.

Th two-body Hermite operator $g$ in Eq.~(\ref{eq:ucom}) is defined as
\begin{eqnarray}
g &=& \frac12 \left\{ p_r s(r)+s(r)p_r\right\} ~,
\label{eq:ucom_g}
\end{eqnarray}
where the operator $p_r$ is the radial component of the relative momentum and is conjugate to the relative coordinate $r$. 
The function $s(r)$ expresses the amount of the shift of the relative wave function at the relative coordinate $r$ for every nucleon pair in the nuclei.  
We use the TOSM basis states to describe $\Phi$.

In UCOM, the function $s(r)$ is determined variationally to minimize the total energy of the system. 
We parametrize $s(r)$ in the same manner as proposed by Feldmeier and Neff \cite{feldmeier98,neff03} for four channels of spin-isospin pairs.
The detailed forms of $s(r)$ and their parametrization are explained in our previous papers on He and Li isotopes \cite{myo11,myo12}.
In the present analysis, we use the same $s(r)$ functions for every state of $^8$Be.
To simplify the numerical calculation, we adopt the ordinary UCOM for the central correlation part instead of the $S$-UCOM in this analysis,
where $S$-UCOM introduces the partial-wave dependence in $s(r)$, in particular between the $s$-wave and other partial waves, to improve the variational solutions \cite{myo09}.
It has been shown that $S$-UCOM improves the short-range part of the relative $d$-wave components of the nucleon pair in nuclei 
and the $sd$ coupling caused by the tensor force can be increased by using $S$-UCOM rather than UCOM.

\section{Results}\label{sec:result}

\subsection{$^4$He}
First, we explain in detail the results of $^4$He, which are important in discussing the structures of $^{8}$Be.
In Table \ref{tab:4He_ham}, we show the results for $^4$He using the AV8$^\prime$ interaction, which consists of central, $LS$ and tensor terms and is used in the benchmark calculation given by Kamada et al., where the Coulomb term is ignored \cite{kamada01}.  
We compare the results between TOSM+UCOM and TOSM+$S$-UCOM. 
When we apply $S$-UCOM instead of the ordinary UCOM, it is found that the energy gain is about 2 MeV in total, which gets closer to the rigorous value \cite{myo09}.  In particular, the contribution from the tensor force becomes large due to the improvement of the $sd$ coupling of the tensor matrix elements, as mentioned earlier. The kinetic energy is increased due to the tensor force properties. 
The total energy of $^4$He is obtained as $-22.30$ MeV in TOSM. The matter radius is obtained as 1.52 fm and the $(0s)^4$ configuration is dominant at 84.1\%. 

\begin{table}[t]
\centering
\caption{Hamiltonian components for $^4$He in MeV.}
\begin{tabular}{c|ccccc}
\noalign{\hrule height 0.5pt}
                   &  Energy  &  Kinetic &  Central & Tensor    &  $LS$   \\
\noalign{\hrule height 0.5pt}
TOSM+UCOM          & $-20.46$ & $ 86.95$ & $-54.63$ & $-51.06$  & $-1.73$ \\   
TOSM+$S$-UCOM      & $-22.30$ & $ 90.50$ & $-55.71$ & $-54.55$  & $-2.53$ \\
SVM~\cite{kamada01}& $-25.92$ & $102.35$ & $-55.23$ & $-68.32$  & $-4.71$ \\
\noalign{\hrule height 0.5pt}
\end{tabular}
\label{tab:4He_ham}
\end{table}

\begin{table}[t]
\centering
\caption{Hamiltonian components in TOFM for $^4$He in MeV.}
\begin{tabular}{c|ccccc}
\noalign{\hrule height 0.5pt}
                   &  Energy  &  Kinetic &  Central & Tensor    &  $LS$   \\
\noalign{\hrule height 0.5pt}
TOFM\cite{horii12} & $-24.18$ & $ 95.50$ & $-54.67$ & $-61.32$  & $-4.09$ \\   
double $Y_2$       & $-25.71$ & $100.89$ & $-54.73$ & $-67.21$  & $-4.66$ \\
\noalign{\hrule height 0.5pt}
\end{tabular}
\label{tab:4He_ham_TOFM}
\end{table}

The Hamiltonian components of $^4$He are compared with the stochastic variational method (SVM) using correlated Gaussian basis functions \cite{varga95,suzuki08}, 
which is one of the rigorous calculations.
These quantities are useful when we discuss $^8$Be, in particular, the $\alpha$ clustering aspect of $^8$Be. 
In TOSM, it is found that the kinetic, tensor, and $LS$ terms give smaller contributions than those in SVM. 
This shortage is considered to come from the contributions of the higher configurations beyond the 2p2h space,  
which is related to the harmonic oscillator assumption of the hole states in TOSM.
The other reason is the two-body approximation of UCOM in the unitary transformation of the Hamiltonian of the $A$-body system. 

As for the comparison with the rigorous calculation, we see that the central contribution in TOSM+UCOM/$S$-UCOM satisfies the rigorous value. 
On the other hand, the tensor and kinetic contributions show some shortage from the rigorous values even in the $S$-UCOM case.  
One of the possibilities explaining the shortage of these values in TOSM is the treatment of the short-range part of the tensor correlations.  
Although the dominant part of the tensor force is of intermediate and long ranges, there remains a small strength in the short-range part of the tensor force, which can couple with the short-range correlations.  

Horii et al. estimate the amount of this coupling by using the variational few-body method of SVM \cite{horii12},
in which the short-range repulsion and the tensor force are directly treated without an approximation such as UCOM.
In Ref.\,\cite{horii12}, they propose the one-pair approximation of the tensor coupling using the single $Y_2$ function,
which introduces the $d$-wave component into the few-body wave function, called thes tensor-optimized few-body model (TOFM). 
It is shown in Table \ref{tab:4He_ham_TOFM} that TOFM gives a good binding energy of $^4$He of 24.18 MeV with AV8$^\prime$ as compared with the rigorous calculation. 
The physical concept of TOFM is the same as that of TOSM except for the use of UCOM. 
These results imply that the UCOM transformation for short-range correlation gives an energy loss of about 1.9 MeV in $^4$He in comparison with the TOFM result for the tensor and kinetic contributions in particular, and also the $LS$ contribution. 
The remaining shortage of the binding energy with respect to the rigorous calculation shown in Table \ref{tab:4He_ham} should come from higher configurations beyond 2p2h excitations in TOSM.  
Horii et al. have confirmed that the two-pair tensor coupling using double $Y_2$ functions in $^4$He by extending TOFM, called double $Y_2$, reproduces the rigorous results within 200 keV \cite{horii12}, which corresponds to the 4p4h mixing in TOSM.

\begin{table}[t]
\centering
\caption{Energies of $^4$He using AV8$^\prime$ in units of MeV by increasing the tensor matrix elements.}
\begin{tabular}{ccccccccc}
\noalign{\hrule height 0.5pt}
$X_{\rm T}$ & 1.0     &  1.1     &   1.2    & 1.3  \\
\noalign{\hrule height 0.5pt}                                     
Energy    & $-20.46$  & $-25.32$ & $-31.13$ & $-37.41$  \\
Kinetic   & $ 86.95$  & $ 93.24$ & $ 99.99$ & $106.46$  \\
Central   & $-54.63$  & $-55.98$ & $-56.90$ & $-57.57$  \\
Tensor    & $-51.06$  & $-60.71$ & $-71.97$ & $-83.66$  \\
LS        & $ -1.73$  & $ -1.87$ & $ -2.25$ & $ -2.64$  \\
\noalign{\hrule height 0.5pt}
\end{tabular}
\label{4He_tensor}
\centering
\caption{Energies of $^4$He using AV8$^\prime$ in units of MeV by increasing the $LS$ matrix elements.}
\begin{tabular}{ccccccccc}
\noalign{\hrule height 0.5pt}
$X_{LS}$& 1.0        &  1.2     & 1.4      & 1.6   \\
\noalign{\hrule height 0.5pt}                                       
Energy  & $ -20.46$  & $-20.34$ & $-20.70$ & $-21.08$   \\
Kinetic & $  86.95$  & $ 87.16$ & $ 88.18$ & $ 89.28$   \\
Central & $ -54.63$  & $-54.87$ & $-54.99$ & $-55.10$   \\
Tensor  & $ -51.06$  & $-50.68$ & $-51.43$ & $-52.23$   \\
LS      & $  -1.73$  & $ -1.95$ & $ -2.45$ & $ -3.02$   \\
\noalign{\hrule height 0.5pt}
\end{tabular}
\label{4He_LS}
\end{table}

Considering the difference between the results of TOSM and TOFM, the three-body UCOM term is one of the possibilities to overcome the lack of energy from UCOM~\cite{myo09, feldmeier98}.  This three-body term may contribute to increasing the tensor energy and introducing more high-momentum components into the wave function. 
Another possibility is the improvement of the correlation function $s(r)$. The functional forms of $s(r)$ are introduced from the consideration of the short-range behavior of the two-body system only with the central $NN$ interaction \cite{feldmeier98}.  It would be interesting to determine an appropriate forms of $s(r)$ in the bare Hamiltonian case, including the tensor force explicitly.

We therefore consider therefore the shortages of the tensor and $LS$ contributions in TOSM.
For the treatment of the tensor correlation in TOSM, we do not use $S$-UCOM \cite{myo09}. 
It was found that the $S$-UCOM gains the tensor contribution more than the ordinary UCOM, which corresponds to about 5\% of the enhancement of the tensor matrix elements.
For the treatment of the $LS$ force, it is found that TOSM underestimates the contribution of the $LS$ force in $^4$He from the rigorous value, as shown in Table \ref{tab:4He_ham}.
One of the reasons for this comes from the short-range nature of the $LS$ force in the AV8$^\prime$ interaction.
The short-range part of $LS$ force can couple with the UCOM transformation, which results in the many-body term: at least the three-body term of the UCOM operator \cite{feldmeier98}.
However, beyond the two-body term, the many-body term beyond two-body one is generally missing in UCOM so far. 
To overcome these properties of the tensor and $LS$ forces in the usage of UCOM, 
we should enhance the corresponding matrix elements and define TOSM with this enhancement of matrix elements toward the calculations of heavier nuclei.

In Tables \ref{4He_tensor} and \ref{4He_LS}, the energies and the Hamiltonian components of $^4$He are shown by increasing the tensor and $LS$ forces, respectively,
where $X_{\rm T}$ and $X_{LS}$ correspond to the enhancement factors of the matrix elements of the tensor and $LS$ forces, respectively.
The factors $X_{\rm T}$=1 and $X_{LS}$=1 indicate the original results.
For the tensor part shown in Table \ref{4He_tensor}, as the tensor contribution increases, the kinetic energy is greatly enhanced.
This is an indication of the high-momentum nature of the tensor force. 
The total energy is also greatly affected by the tensor contribution.
The $LS$ contribution is slightly enhanced because the tensor force generally produces the particle-hole excitations in the configuration
that affect the $LS$ contribution. The central contribution does not change so much and this force is independent of the tensor correlation.
In comparison with the TOFM results shown in Table \ref{tab:4He_ham_TOFM}, the tensor contribution of about $-61$ MeV can be simulated with $X_{\rm T}$=1.1 to get the full contribution of the 2p2h configurations.

For the $LS$ part shown in Table \ref{4He_LS}, the same trend seen in the tensor case is confirmed, however, the increase of each component is not as drastic as the tensor case in Table \ref{4He_tensor}
and the tensor contribution does not change so much.
Only the $LS$ contribution is increased using the enhancement factor $X_{LS}$.

\begin{table}[tb]
\centering
\caption{Hamiltonian components for $^4$He in MeV using TOSM with the ($X_{\rm T}$, $X_{LS}$)=(1.1, 1.4) set.}
\begin{tabular}{ccccc}
\noalign{\hrule height 0.5pt}
  Energy  &  Kinetic &  Central & Tensor    &  $LS$   \\
\noalign{\hrule height 0.5pt}
 $-26.16$ & $ 95.45$ & $-56.17$ & $-62.43$  & $-3.02$ \\
\noalign{\hrule height 0.5pt}
\end{tabular}
\label{tab:4He_ham2}
\end{table}

\begin{table}[t]  
\centering
\caption{Occupation numbers in each orbit of $^4$He using AV8$^\prime_{\rm eff}$ and Minnesota (MN) interactions.}
\begin{tabular}{c|cccccccc}
\noalign{\hrule height 0.5pt}
$^4$He$(J^\pi)$  &~$0s_{1/2}$~&~$p_{1/2}$~&~$p_{3/2}$~&~$1s_{1/2}$~&~$d_{3/2}$~&~$d_{5/2}$~\\ 
\noalign{\hrule height 0.5pt}                                                 
AV8$^\prime_{\rm eff}$ & ~3.72~  &~ 0.07 ~   & ~ 0.05 ~  &~ 0.05 ~    &~ 0.04  ~  &~ 0.02~    \\ 
MN                     & ~3.94~  &~ 0.01 ~   & ~ 0.03 ~  &~ 0.01 ~    &~ 0.004 ~  &~ 0.006~   \\ 
\noalign{\hrule height 0.5pt}
\end{tabular}
\label{occ4}
\end{table}

Considering the different roles of the tensor and $LS$ forces on the structure of $^4$He, 
we try to include these effects in TOSM to simulate phenomenologically the TOFM results of $^4$He as closely as possible by increasing the corresponding matrix elements.
We enhance the tensor matrix elements by 10\%, namely, $X_{\rm T}$=1.1 and the $LS$ matrix elements by 40\% as $X_{LS}$=1.4, respectively. 
The Hamiltonian components of $^4$He in the revised TOSM calculations are listed in Table \ref{tab:4He_ham2}.
It is found that the total energy is good and more than 90\% of the tensor component and kinetic energy of the benchmark calculation (SVM) shown in Table \ref{tab:4He_ham} is reproduced.
It is also found that the present results of $^4$He almost reproduce the TOFM solutions.
This indicates that the missing effects of TOSM are effectively recovered by using the phenomenological enhancements of the matrix elements.
Hereafter, we use this parameter set as the effective interaction of TOSM and call this interaction ``AV8$^\prime_{\rm eff}$'', which retains the characteristics of the bare $NN$ interaction.
We use TOSM with the use of AV8$^\prime_{\rm eff}$ for the analysis of the structures of $^8$Be.

To see the effect of the tensor force in TOSM, we show the results with the effective Minnesota (MN) $NN$ interaction, which does not have the tensor force.
We choose the $u$ parameter as 0.95 for the central force and use set III of the $LS$ force \cite{reichstein70, tang78}.
Using the MN interaction, the binding energy of $^4$He is obtained as 30.55 MeV without the Coulomb interaction. 
The radius of $^4$He is obtained as 1.39 fm, which is smaller than the experimental value and also the value calculated in TOSM (1.52 fm).
The $(0s)^4$ configuration is quite dominant in $^4$He by 96.6\%.

We compare the occupation numbers of nucleons in $^4$He between AV8$^\prime_{\rm eff}$ and MN interactions in Table \ref{occ4}.
In AV8$^\prime_{\rm eff}$, it is shown that the $p_{1/2}$ orbit has the largest contribution among the particle states according to the large 2p2h mixing.
In the MN case, it is found that the component of the $0s$ orbit is larger than the AV8$^\prime_{\rm eff}$ case and the enhancement of the $p_{1/2}$ orbit is not obtained.
These results mean that the tensor force introduces the specific excitations from the $s$-shell to the $p$- and $sd$-shells in $^4$He,
which is related to the properties of the tensor operator $S_{12}$ \cite{myo07,myo09}. 
In the next section of the analysis of $^8$Be, we also compare the results between two interaction cases.

\subsection{$^8$Be}\label{sec:Be}

\subsubsection{Energy spectrum}\label{sec:Be_spectrum}

We discuss the structures of $^8$Be in TOSM using AV8$^\prime_{\rm eff}$ determined from the $^4$He analysis. 
The total binding energy of the ground state of $^8$Be is obtained as $30.19$ MeV, which is smaller than the experimental value $56.50$ MeV; we discuss this point later. 
We show the excitation energy spectra of $^8$Be in Fig.\,\ref{fig:AV_Be8}.
It is found that there are two groups of states in the spectrum; one is the three ground band states of $0^+_1$, $2^+_1$, and $4^+_1$ and the other is the highly excited states above $4^+_1$.  
The relative energy of the two groups in $^8$Be is about 5 MeV in the experiments between $4^+_1$ and $2^+_2$, while it is about 2 MeV in the calculated spectrum.
The three ground band states of $0^+_1$, $2^+_1$ and $4^+_1$ seem to form a rotational band and the energy splitting is slightly narrower than the experiment.
From the spectrum, the lowest three states are considered as deformed (clustered) rotational states.
It is found that the highly excited states include the $T=1$ states starting from $2^+$. 
This state is the isobaric analog state of the $^8$Li ground state and the energy spectra of Li isotopes have been successfully described using TOSM in the previous study \cite{myo12}.
In the experiment, the $2^+(T=1)$ state is almost degenerate with the $2^+(T=0)$ state in energy and this situation is nicely reproduced in TOSM.

\begin{figure}[t]
\centering
\includegraphics[width=6.0cm,clip]{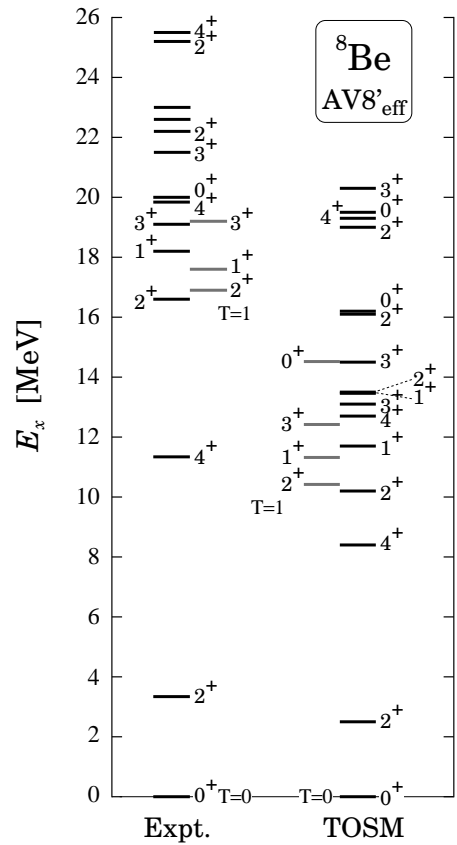}
\caption{Excitation energy spectrum of $^8$Be using AV8$^\prime_{\rm eff}$.
Shown on the left-hand side are experimental data for $T$=0 and $T$=1 (gray lines), listed separately.  
Shown on the right-hand side are theoretical results for $T$=0 and $T$=1, again listed separately.}
\label{fig:AV_Be8}
\end{figure}

\begin{figure}[tbh]
\centering
\includegraphics[width=6.0cm,clip]{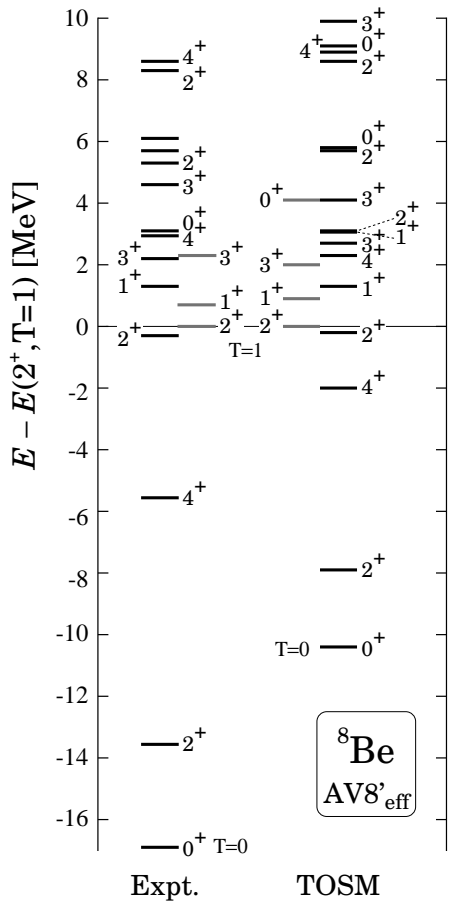}
\caption{Energy spectrum of $^8$Be using TOSM normalized to the $2^+$($T$=1) state.}
\label{fig:AV_Be8_2}
\end{figure}

For the ground band states of $^8$Be, 
these states have been studied and are understood as two-$\alpha$ clustering states by using the $\alpha$ cluster model in the low excitation energy region \cite{horiuchi12,yamamoto10}.
The matter radius of the $^8$Be ground state is obtained as 2.21 fm in TOSM, which is smaller than the value of Brink's two-$\alpha$ cluster model of 2.48 fm in the bound state approximation, 
discussed again later.
The small radius in TOSM indicates that the $\alpha$ clustering is not fully described in the present solution of $^8$Be.
We consider possible reasons for this as follows:
From the viewpoint of the shell model, it is generally difficult to express the asymptotic form of the spatially developed $\alpha$ clustering states that contain high shell quanta components.
In TOSM, it is found that specific 2p2h excitations involving high-momentum components are essential to incorporate the tensor correlation in the single $\alpha$ particle \cite{myo09,myo11}.
This fact indicates that when two-$\alpha$ particles are established in $^8$Be, each $\alpha$ particle independently needs the 2p2h components to express the tensor correlation.
The other possibility is that many kinds of particle-hole excitations in the shell model bases are also important to describe the formation of the separated two-$\alpha$ clusters in space.
On the other hand, we take up to 2p2h excitations in the present TOSM and this approximation restricts the spatial cluster formation in $^8$Be, which can require the 4p4h excitations for the $\alpha$ clustering.
In addition, the tensor correlations in these $\alpha$ particles might require higher excitations.
Considering the small relative energy between the three ground band states and the highly excited states shown in Fig.\,\ref{fig:AV_Be8_2}, 
the three ground band states are desired to further increase the energy by extending the model space of TOSM to represent the $\alpha$ clustering correlation.
Later we estimate the correlation energy of the $\alpha$ clustering explicitly using the cluster model.
It is interesting to extend the space of TOSM so as to express the $\alpha$ clustering correlation, which includes the tensor correlation in each $\alpha$ particle. 
We intend to study this in the future.

We discuss the highly excited states above $4^+_1$ with the $T$=0 and $T$=1 states shown in Fig.\,\ref{fig:AV_Be8_2}.
To see those states more clearly, we show the energy spectrum normalized to the $2^+$ state with $T$=1 in Fig.\,\ref{fig:AV_Be8_2}.
Considering the TOSM results for Li isotopes, the $T$=1 states can be interpreted as being dominantly the shell-like state, not the clustering state.
Similarly, TOSM is considered to describe the dominant components of the $T$=1 states of $^8$Be on the basis of the shell-model-type configurations.
Hence, it is meaningful to adopt the $2^+$($T$=1) state as a reference state to show the $^8$Be spectrum.

In TOSM, the degeneracy of the $2^+_2$ ($T$=0) and $2^+$ ($T$=1) states is consistent with the experiment.
The level spacings are reproduced fairly well in the $T$=0 and $T$=1 states, except for some levels such as $0^+_2$ ($T$=0).
The degeneracy of the $T$=0 and $T$=1 states will be discussed later in connection with the tensor and $LS$ forces in the $NN$ interaction.
For $T$=0 states staring from $2^+_2$, it is found that the overall spectrum almost agrees with the experiment including the level ordering.
The energy spacing between $2^+_2$ and $1^+_1$ is consistent with the experiment. 
At 8$-$10 MeV, we obtain four states, some of which are consistent with the experimental states of $2^+_4$ and $4^+_3$.
For the $T$=1 states, the order of the states and their relative energies are almost reproduced. The $0^+$($T$=1) state has not yet been confirmed experimentally.

The TOSM with AV8$'_{\rm eff}$ is found to reproduce well the excitation energy spectrum of $^8$Be in the higher energy region in particular.
It is interesting to examine the roles of the non-central tensor and $LS$ matrix elements in determining the energy spectrum of $^8$Be.
In particular, the tensor force is the most important ingredient for $^8$Be, similar to $^4$He. 
For this purpose, we try to see the structures of $^8$Be by changing the strengths of two forces around AV8$^\prime_{\rm eff}$.
The same analysis was performed for $^4$He \cite{myo05} and it was found that the specific particle-hole excitation from the $s_{1/2}$ to $p_{1/2}$ orbits with the same $j$-value is 
very sensitive to the tensor force.
This result can be understood from the properties of the tensor operator $S_{12}$ with the rank of two for the orbital angular momentum and intrinsic spin parts, respectively.
The specific excitation caused by the tensor force is important for understanding the Pauli-blocking effect on He isotopes \cite{myo11}, Li isotopes \cite{myo12} and also on the halo formation in $^{11}$Li \cite{myo07_11}.

\begin{figure*}[t]
\centering
\includegraphics[width=11.0cm,clip]{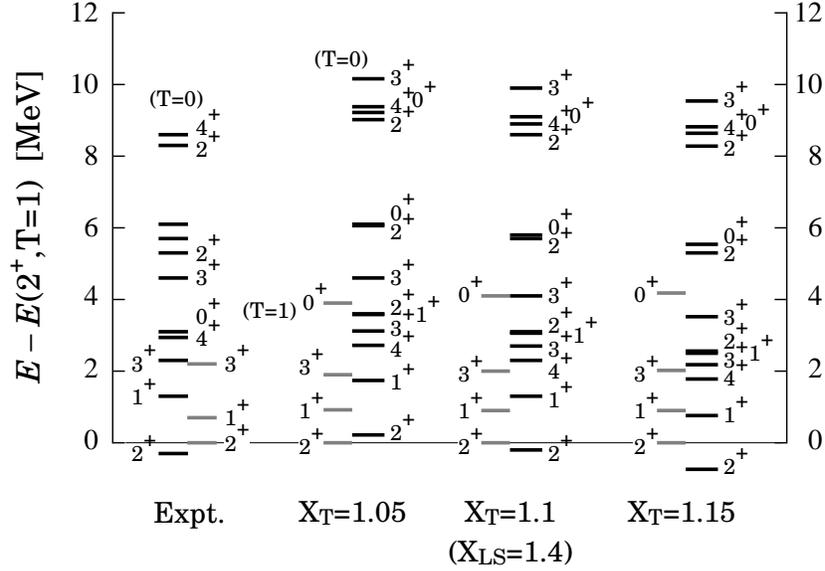}
\caption{The energy spectrum of highly excited states in $^8$Be with TOSM by changing the tensor strength about $\pm 5\%$ around AV8$^\prime_{\rm eff}$ keeping $X_{LS}=1.4$.  
The energy spectrum is normalized to the $2^+$($T$=1) state, where the black lines denote the $T=$0 states and the gray lines the $T$=1 states.} 
\label{fig:AV_Be8_tensor2}
\end{figure*}

We want to study the effect of the tensor and spin-orbit forces on the spectrum of the highly excited states in $^{8}$Be.
We start from the effective interaction AV8$^\prime_{\rm eff}$, namely, $X_{\rm T}$=1.1 and $X_{LS}$=1.4 and change one of them around these values, individually. 
We have confirmed that the ground band structures, such as the energy spacings do not depend on the changes of $X_{\rm T}$ and $X_{LS}$ in this parameter range.
On the other hand, highly excited states are somewhat influenced; the results of this are shown in Fig.\,\ref{fig:AV_Be8_tensor2} for the tensor force case, and in Fig.\,\ref{fig:AV_Be8_LS2} for the $LS$ force case.
In Figs.\,\ref{fig:AV_Be8_tensor2} and \ref{fig:AV_Be8_LS2}, we normalize the spectrum of $^8$Be to the $2^+$($T$=1) state.

The highly excited states of $^8$Be are shown in Fig.\,\ref{fig:AV_Be8_tensor2}, where the tensor strength is changed about $\pm 5\%$ from AV8$'_{\rm eff}$.
It is clearly seen that the energies of the $T$=0 states go down with respect to the $T$=1 states. 
The reason for this trend is considered as follows; the $T$=0 states are more sensitive to the tensor correlation than the $T$=1 states because of the strong attractive nature
of the $T$=0 channel of the tensor force. Hence, as the tensor correlation becomes stronger, the $T$=0 states gain more energy than the $T$=1 states.
Essentially, the level orders in each $T$=0 and $T$=1 state do not change so much. The Hamiltonian components of each state are discussed in the next subsection.
The same effect of tensor force on the isospin of nuclei is expected in other $N=Z$ nuclei, such as $^4$He and $^6$Li \cite{myo12}.
For $^6$Li case, $T=0$ and $T=1$ states coexist in the low excitation energy region and it is interesting to investigate the effect of the tensor force in those states.

For the change in the $LS$ force shown in Fig.\,\ref{fig:AV_Be8_LS2}, the energy spectrum changes slightly with respect to the enhancement of the $LS$ matrix elements.  For highly excited states, the energy spectrum becomes wider. 
For example, the energy spacing between $1^+_1$ and $2^+_2$ in $T$=0 state becomes larger as the $LS$ matrix elements increase.
The $4^+_2$ state also comes down energetically.
The four states above 8 MeV in Fig.\,\ref{fig:AV_Be8_LS2} go up in energy and the level order of $2^+ - 4^+$ is kept, which is consistent with the experiment.
The $T$=1 states are also getting sparse and the level order does not change. 
From these $^8$Be results, it is found that the tensor force affects the relative energies between the $T$=0 and $T$=1 states and that the $LS$ force affects the spacings of each energy level. 

\begin{figure*}[t]
\centering
\includegraphics[width=11.0cm,clip]{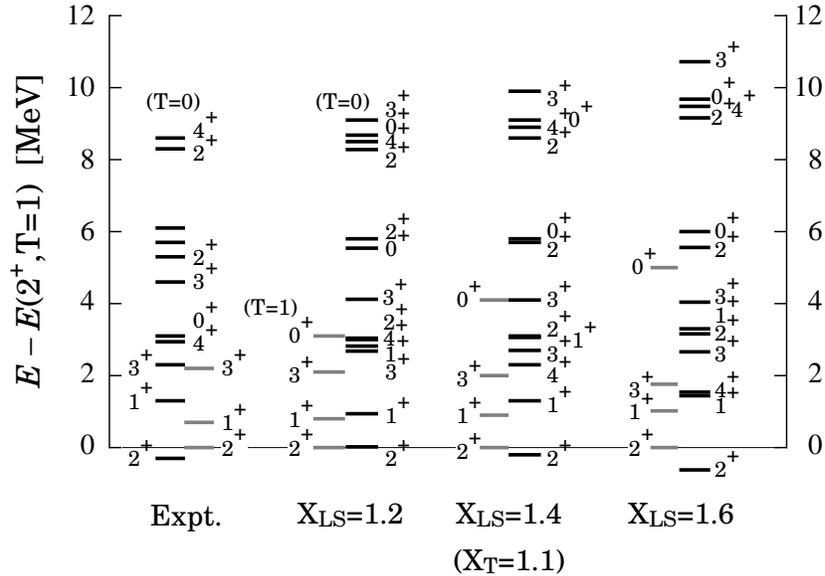}
\caption{The energy spectrum of highly excited states in $^8$Be with TOSM by changing the spin-orbit strength about $\Delta X_{LS}=\pm 0.2$ around AV8$^\prime_{\rm eff}$ keeping $X_{\rm T}=1.1$.  
The energy spectrum is normalized to the $2^+$($T$=1) state, where the black lines denote the $T=$0 states and the gray lines the $T$=1 states.} 
\label{fig:AV_Be8_LS2}
\end{figure*}

We have already discussed the effect of $\alpha$ clustering in improving the ground band states of $^8$Be.
It is important to estimate the energy gain due to the inclusion of the two-$\alpha$ clustering component in $^8$Be.
For this purpose, we describe $^8$Be in terms of Brink's two $\alpha$ cluster model assuming the $(0s)^4$ configuration for each $\alpha$ particle.
We use the effective interaction of Volkov No.2 \cite{volkov65} with a Majorana parameter as 0.6.
The size of the $\alpha$ particle is chosen to reproduce the observed radius of 1.5 fm.
In Fig.\,\ref{fig:brink}, the calculated energy surfaces of the three ground band states are shown as functions of the matter radius of $^8$Be. 
We also perform the calculation of generator coordinate method (GCM) using the basis states on each energy surface.
The ground band structure is well reproduced in the GCM results and the ground $0^+$ state is located just above the two-$\alpha$ threshold energy by about 300 keV.
This is consistent with Ikeda's threshold rule of the clustering state \cite{ikeda68}.
The radius of the ground state in GCM is 2.76 fm.
To estimate the correlation energy coming from the $\alpha$ clustering component, we refer to the radius of $^8$Be in TOSM as 2.21 fm.
The ground state energy at this radius is $-51.5$ MeV and the GCM result is $-55.5$ MeV in Fig.\,\ref{fig:brink}. 
The difference between the two energies, 4 MeV, can naively correspond to the correlation energy by the inclusion of the $\alpha$ clustering component in $^8$Be.
Hence, the inclusion of the $\alpha$ clustering component in TOSM can improve the energy spectrum in TOSM shown in Fig.\,\ref{fig:AV_Be8_2}.
In particular, the small energy spacing between the ground band states and the highly excited states is expected to be recovered.
It would be interesting to include the $\alpha$ clustering basis states in TOSM explicitly to describe the $^8$Be nucleus from the ground states to the highly excited states.

\begin{figure}[t]
\centering
\includegraphics[width=8.0cm,clip]{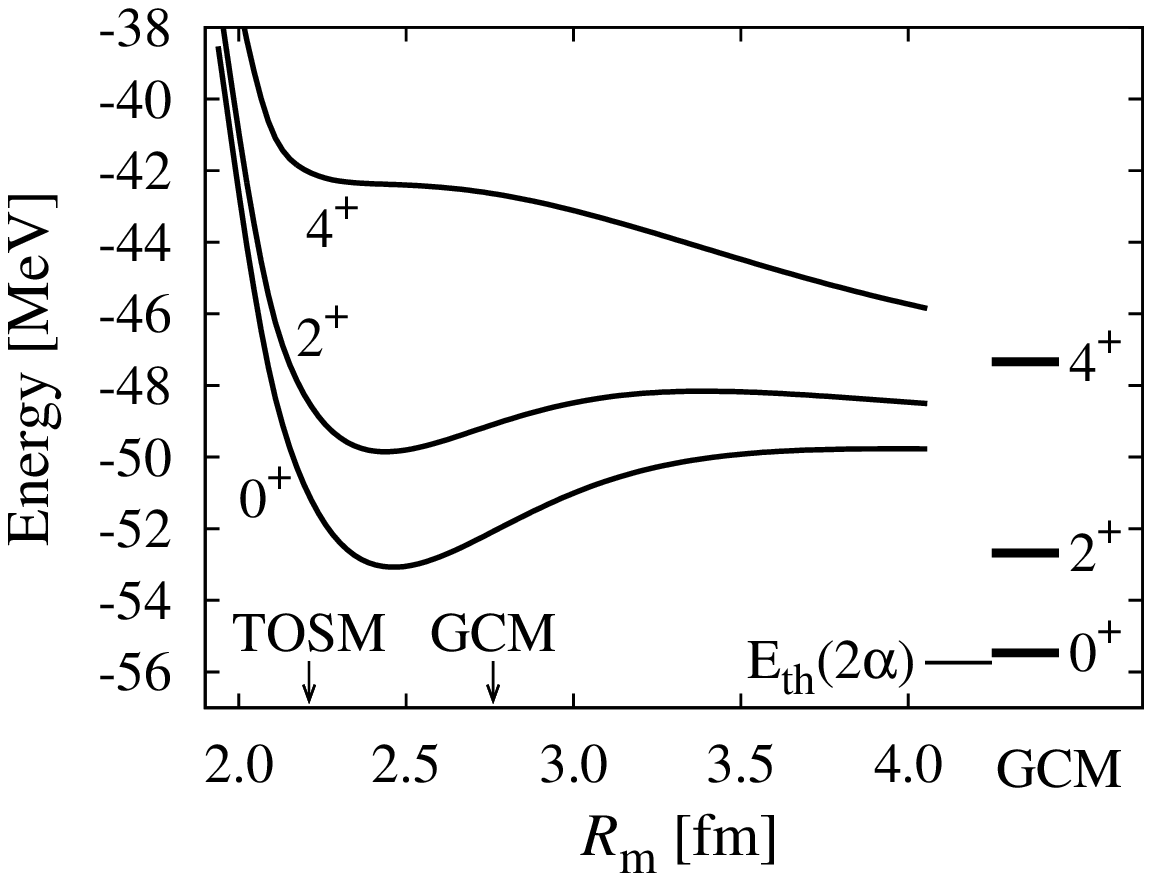}
\caption{Energy surfaces of $^8$Be using Brink's two-$\alpha$ cluster model with respect to the matter radius, $R_{\rm m}$. 
The GCM results are shown in the right-hand part with a threshold energy of two $\alpha$.
The radii obtained using TOSM and GCM are shown by the arrows.}
\label{fig:brink}
\end{figure}

\begin{figure}[t]
\centering
\includegraphics[width=6.0cm,clip]{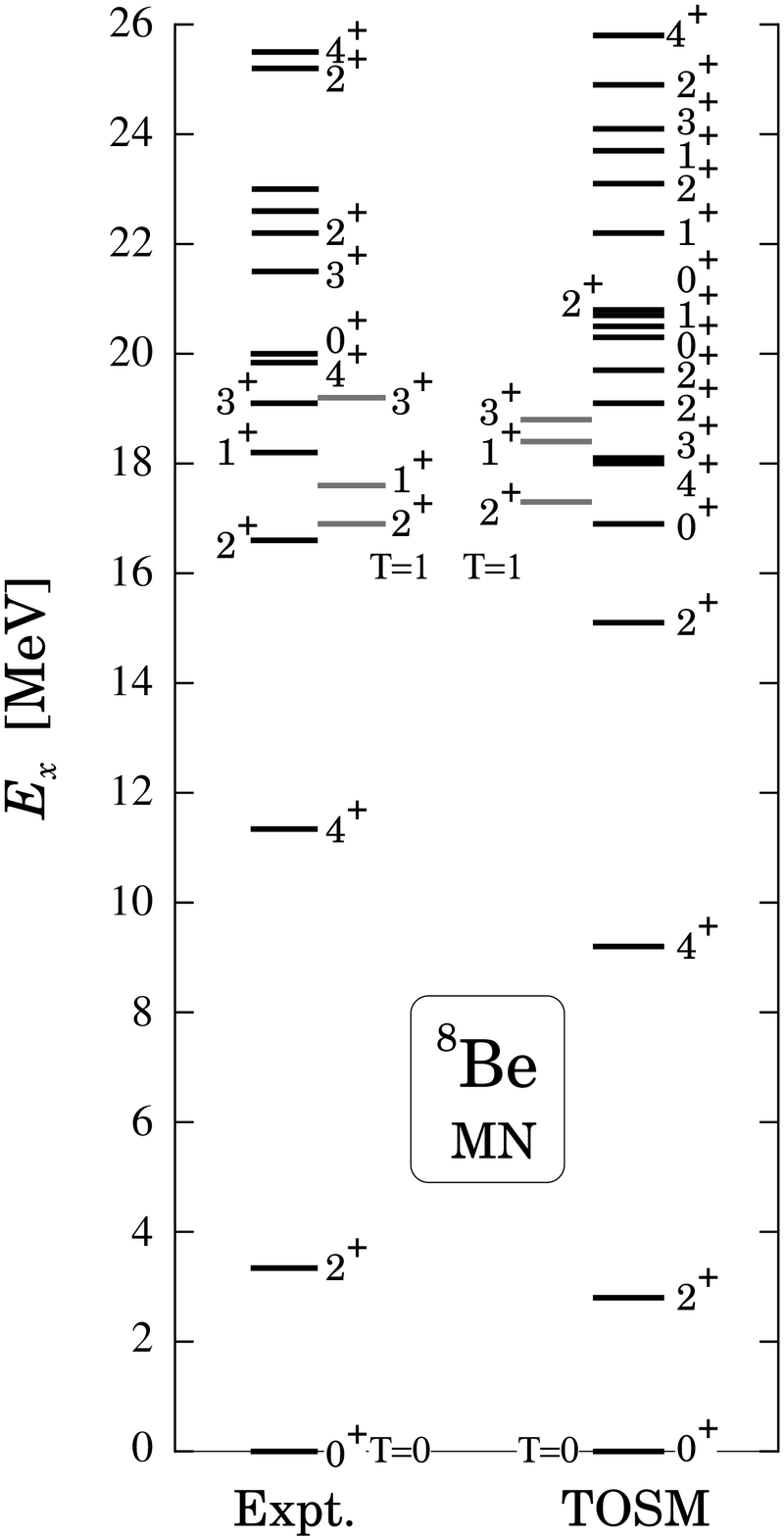}
\caption{Excitation energy spectrum of $^8$Be with TOSM using the Minnesota interaction (MN).}
\label{fig:MN_Be8}
\end{figure}

We compare the energy spectrum of $^8$Be with those using the effective MN interaction to find the interaction dependence and in particular the effect of the tensor force.
In MN, it has been discussed that the $LS$ force gives a stronger effect than the AV8$^\prime$ case to describe the $LS$ splitting energies \cite{myo11,myo12}, 
because MN was originally constructed within the restricted configuration space of the nucleus such as the $(0s)^4$ assumption of $^4$He, 
and the tensor correlation is renormalized into the effective interaction and the model space.  
In this study, we reduce the strength of the $LS$ force by 30\% to give the same $LS$ splitting energy, 1.5 MeV, as $^5$He, obtained using AV8$^\prime$ in TOSM \cite{myo11}.
The binding energy of the $^8$Be ground state is obtained as $60.98$ MeV, which is close to the experimental value of $56.50$ MeV.
This interaction gives a radius of 1.89 fm for $^8$Be smaller than the AV8$^\prime$ case of 2.21 fm and similar to the $^4$He result, obtained as 1.39 fm.
This means that MN cannot reproduce the nuclear saturation properties, which is closely related to the role of the tensor force. 

We discuss the excitation energy spectrum using MN shown in Fig.\,\ref{fig:MN_Be8}, which reproduces the overall excitation energies.
At lower excitation energies, we obtain three ground band states and the rotational structure is nicely reproduced. 
Above 14 MeV excitation energy, many spin states are obtained.
For $T$=0 states, the relative energy between the $4^+_1$ and $2^+_2$ states is fairly well reproduced, which is different from the AV8$^\prime_{\rm eff}$ results shown in Fig.\,\ref{fig:AV_Be8}.
In MN, the effect of the tensor force is renormalized into the model space, hence the 2p2h excitation induced by the tensor force is not necessary to describe $^4$He.
As a result, some amount of two-$\alpha$ clustering component in $^8$Be can be described in TOSM in terms of the 2p2h excitations,
which could be the reason for the good energy spacing between $4^+_1$ and $2^+_2$.
This situation is different from the case of AV8$^\prime_{\rm eff}$ with the tensor force.

For highly excited states, the $T=0$ states are located lower in energy than the $T=1$ states by about 2 MeV.
This comes from the strong $T=0$ channel of the central force in MN.
In $T$=0, it is found that the level order is often different from the experiment.
For example, the $0^+_2$ state is obtained much lower than the experiment and the $1^+_1$ state is located much higher than the experiment.
The order between the $1^+_1$ and $3^+_1$ states is also opposite to that seen in the experiment.
From the results, the level order of the higher excited states are rather better described in AV8$^\prime_{\rm eff}$ in Fig.\,\ref{fig:AV_Be8} than the MN case.
This difference is related to the tensor force that is missing in MN explicitly.
The tensor force can give the correct state dependence to reproduce the excitation energy spectrum of $^8$Be.
The similar discussion of the tensor force effect in comparison with  MN has been presented for the excited states of $^4$He \cite{horiuchi13}.
In Fig. \ref{fig:MN_Be8}, for $T$=1 with MN, the excitation energies of the $T$=1 states almost agree with the experiment and the energy spacing is slightly smaller than the experiments.
The relative energy of the $2^+_2$ ($T$=0) state and the $2^+$ ($T$=1) state are larger than the experiments, in which the $2^+_2$ ($T$=0) state is lower.

\subsubsection{Hamiltonian components}\label{sec:Be_ham}

We discuss the Hamiltonian components in each state of $^8$Be in TOSM to discuss the importance of the tensor force.
In Table \ref{tab:8Be_ham}, we show the Hamiltonian components of the $^8$Be states.

\begin{table}[t]
\centering
\caption{Hamiltonian components in MeV for $^8$Be.}
\begin{tabular}{c|ccccc}
\noalign{\hrule height 0.5pt}
State ($T$=0)   &  Energy  &  Kinetic &  Central & Tensor    &  $LS$   \\
\noalign{\hrule height 0.5pt}
$0^+_1$   & $-30.19$ & $ 192.43$ & $-115.33$ & $-96.52$  & $-10.77$ \\
$0^+_2$   & $-13.96$ & $ 181.51$ & $-96.84$  & $-87.43$  & $-11.20$ \\
$0^+_3$   & $-10.65$ & $ 177.84$ & $-91.77$  & $-87.75$  & $- 8.98$ \\
\noalign{\hrule height 0.5pt}
$1^+_1$   & $-18.51$ & $ 183.29$ & $-95.52$  & $-92.23$  & $-10.05$ \\
$1^+_2$   & $-16.72$ & $ 183.27$ & $-95.70$  & $-93.44$  & $-10.86$ \\
$1^+_3$   &  $-6.68$ & $ 175.26$ & $-88.01$  & $-87.34$  & $- 6.59$ \\
\noalign{\hrule height 0.5pt}
$2^+_1$   & $-27.73$ & $ 191.23$ & $-112.30$ & $-95.24$  & $-11.43$ \\
$2^+_2$   & $-20.02$ & $ 185.36$ & $-98.34$  & $-92.15$  & $-14.88$ \\
$2^+_3$   & $-16.69$ & $ 182.80$ & $-94.54$  & $-93.02$  & $-11.93$ \\
$2^+_4$   & $-14.08$ & $ 181.43$ & $-92.50$  & $-89.78$  & $-13.23$ \\
$2^+_5$   & $-11.11$ & $ 184.16$ & $-90.79$  & $-91.25$  & $-16.33$ \\
\noalign{\hrule height 0.5pt}
$3^+_1$   & $-17.11$ & $ 183.77$ & $-96.32$  & $-92.03$  & $-12.53$ \\
$3^+_2$   & $-15.71$ & $ 182.29$ & $-91.92$  & $-93.30$  & $-12.78$ \\
$3^+_3$   & $- 9.92$ & $ 177.70$ & $-92.55$  & $-88.31$  & $- 6.77$ \\
\noalign{\hrule height 0.5pt}
$4^+_1$   & $-21.77$ & $ 188.46$ & $-105.56$ & $-91.71$  & $-12.96$ \\
$4^+_2$   & $-17.51$ & $ 184.38$ & $-92.22$  & $-92.09$  & $-17.59$ \\
$4^+_3$   & $-10.84$ & $ 179.40$ & $-93.53$  & $-87.70$  & $- 9.00$ \\
\noalign{\hrule height 0.5pt}
\end{tabular}

\begin{tabular}{c|ccccc}
\noalign{\hrule height 0.5pt}
State ($T$=1)  &  Energy  &  Kinetic  &  Central  & Tensor    &  $LS$   \\
\noalign{\hrule height 0.5pt}
$0^+$         & $-15.73$ & $ 158.21$ & $-89.32$  & $-78.42$  & $- 6.21$ \\
$1^+$         & $-18.86$ & $ 167.47$ & $-92.81$  & $-81.80$  & $-11.72$ \\
$2^+$         & $-19.77$ & $ 167.95$ & $-93.78$  & $-81.50$  & $-12.43$ \\
$3^+$         & $-17.83$ & $ 167.44$ & $-91.67$  & $-80.03$  & $-13.58$ \\
\noalign{\hrule height 0.5pt}
\end{tabular}
\label{tab:8Be_ham}
\end{table}

It is found that the ground state possesses the largest tensor contribution and also the largest kinetic energy and central contributions.
For this state, the central contribution of $-115$ MeV is very close to twice the $^4$He value of $-56$ MeV as shown in Table \ref{tab:4He_ham2}.
The kinetic energy of 192 MeV is also very close to twice the $^4$He value of $95$ MeV.
These results represent the possibility of the signature of the two-$\alpha$ structure of $^8$Be.
For the tensor contribution, $-97$ MeV is less than the twice of the $^4$He case, $-62$ MeV.
As has already been discussed, from the viewpoint of the $\alpha$ clustering, the tensor contribution in $^8$Be is favored to have twice the $^4$He value, naively. 
In GFMC \cite{wiringa00}, the contribution of the one-pion exchange potential, the major origin of the tensor force, is calculated for the $^8$Be ground state.
Its contribution is about 2.2 times the $^4$He value, which supports the two-$\alpha$ picture of $^8$Be.
A possible reason for the lack of the tensor contribution is the truncation of the particle-hole excitation in TOSM, as was discussed in the excitation energy spectrum of $^8$Be.
It is interesting to extend the model space of TOSM and see the Hamiltonian components of $^8$Be in comparison with the $^4$He case.
For $2^+_1$ and $4^+_1$, members of the ground band, their contributions from the central and tensor forces and kinetic energies are larger than other same spin states.
This tendency suggests the signature of the common internal structure of $^8$Be for the three ground band states.

Except for the three ground band states, the highly excited states show similar values of the kinetic energies, central, and tensor components.
The $LS$ contribution depends on the states, because the $LS$ matrix elements generally depend on the single-particle configurations, such as the occupation of the $p_{1/2}$ and $p_{3/2}$ orbits.

\begin{table*}[t]
\centering
\caption{Occupation numbers in each orbit of $^8$Be.}
\begin{tabular}{c|cccccccccc}
\noalign{\hrule height 0.5pt}
State ($T$=0)    &~$0s_{1/2}$~  &~$0p_{1/2}$~&~$0p_{3/2}$~&~$1s_{1/2}$~&~$d_{3/2}$~&~$d_{5/2}$~&~$1p_{1/2}$~&~$1p_{3/2}$~\\ 
\noalign{\hrule height 0.5pt}                                                 
$0^+_1$          & ~3.74~     &~ 0.74 ~   & ~ 3.11 ~  &~ 0.05 ~    &~ 0.06  ~  &~ 0.05~    & ~ 0.03~   & ~0.05~    \\ 
$0^+_2$          & ~3.73~     &~ 0.83 ~   & ~ 3.03 ~  &~ 0.06 ~    &~ 0.06  ~  &~ 0.05~    & ~ 0.03~   & ~0.05~    \\ 
$0^+_3$          & ~3.75~     &~ 1.58 ~   & ~ 2.28 ~  &~ 0.05 ~    &~ 0.06  ~  &~ 0.06~    & ~ 0.03~   & ~0.04~    \\ 
\noalign{\hrule height 0.5pt}
$1^+_1$          & ~3.74~     &~ 1.08 ~   & ~ 2.78 ~  &~ 0.05 ~    &~ 0.06  ~  &~ 0.05~    & ~ 0.03~   & ~0.05~    \\ 
$1^+_2$          & ~3.75~     &~ 1.15 ~   & ~ 2.69 ~  &~ 0.05 ~    &~ 0.06  ~  &~ 0.05~    & ~ 0.03~   & ~0.05~    \\ 
$1^+_3$          & ~3.76~     &~ 1.93 ~   & ~ 1.92 ~  &~ 0.05 ~    &~ 0.06  ~  &~ 0.05~    & ~ 0.03~   & ~0.04~    \\ 
\noalign{\hrule height 0.5pt}
$2^+_1$          & ~3.74~     &~ 0.70 ~   & ~ 3.15 ~  &~ 0.05 ~    &~ 0.06  ~  &~ 0.05~    & ~ 0.03~   & ~0.05~    \\ 
$2^+_2$          & ~3.73~     &~ 0.38 ~   & ~ 3.47 ~  &~ 0.06 ~    &~ 0.06  ~  &~ 0.05~    & ~ 0.03~   & ~0.05~    \\ 
$2^+_3$          & ~3.74~     &~ 0.95 ~   & ~ 2.91 ~  &~ 0.06 ~    &~ 0.06  ~  &~ 0.05~    & ~ 0.03~   & ~0.05~    \\ 
$2^+_4$          & ~3.74~     &~ 0.86 ~   & ~ 2.99 ~  &~ 0.06 ~    &~ 0.06  ~  &~ 0.05~    & ~ 0.03~   & ~0.05~    \\ 
\noalign{\hrule height 0.5pt}
$3^+_1$          & ~3.74~     &~ 0.98 ~   & ~ 2.87 ~  &~ 0.05 ~    &~ 0.06  ~  &~ 0.05~    & ~ 0.03~   & ~0.05~    \\ 
$3^+_2$          & ~3.74~     &~ 0.97 ~   & ~ 2.87 ~  &~ 0.06 ~    &~ 0.06  ~  &~ 0.05~    & ~ 0.03~   & ~0.05~    \\ 
$3^+_3$          & ~3.76~     &~ 1.91 ~   & ~ 1.95 ~  &~ 0.05 ~    &~ 0.06  ~  &~ 0.05~    & ~ 0.03~   & ~0.05~    \\ 
\noalign{\hrule height 0.5pt}
$4^+_1$          & ~3.73~     &~ 0.46 ~   & ~ 3.39 ~  &~ 0.06 ~    &~ 0.06  ~  &~ 0.05~    & ~ 0.03~   & ~0.06~    \\ 
$4^+_2$          & ~3.74~     &~ 0.66 ~   & ~ 3.18 ~  &~ 0.06 ~    &~ 0.06  ~  &~ 0.05~    & ~ 0.03~   & ~0.05~    \\ 
\noalign{\hrule height 0.5pt}
\end{tabular}

\begin{tabular}{c|cccccccccc}
\noalign{\hrule height 0.5pt}
State ($T$=1)    &~$0s_{1/2}$~&~$0p_{1/2}$~&~$0p_{3/2}$~&~$1s_{1/2}$~&~$d_{3/2}$~&~$d_{5/2}$~&~$1p_{1/2}$~&~$1p_{3/2}$~\\ 
\noalign{\hrule height 0.5pt}                                                 
$0^+$            & ~3.74~     &~ 1.55 ~   & ~ 2.31 ~  &~ 0.05 ~    &~ 0.06  ~  &~ 0.05~    & ~ 0.04~   & ~0.07~    \\ 
$1^+$            & ~3.73~     &~ 0.47 ~   & ~ 3.39 ~  &~ 0.06 ~    &~ 0.06  ~  &~ 0.05~    & ~ 0.04~   & ~0.07~    \\ 
$2^+$            & ~3.72~     &~ 0.38 ~   & ~ 3.48 ~  &~ 0.06 ~    &~ 0.06  ~  &~ 0.05~    & ~ 0.04~   & ~0.07~    \\ 
$3^+$            & ~3.72~     &~ 0.20 ~   & ~ 3.66 ~  &~ 0.04 ~    &~ 0.06  ~  &~ 0.05~    & ~ 0.04~   & ~0.07~    \\ 
\noalign{\hrule height 0.5pt}
\end{tabular}
\label{occ8}
\end{table*}

It is interesting to discuss the structure differences between the $T$=0 and $T$=1 states of $^8$Be from the viewpoint of the tensor force.
In Table \ref{tab:8Be_ham}, it is shown that the $T$=1 states possess the smaller tensor contributions than the $T$=0 case.
The kinetic energies also show a similar trend due to the high-momentum component of the tensor force.
This result for the $T$=1 states can be related to the isospin dependence of the tensor force, in which the $T$=0 states involve a stronger tensor correlation than the $T$=1 case, 
although the direct relation between the state and the interaction for the isospin property should be carefully examined.

In Table \ref{occ8}, we list the occupation numbers of the $^8$Be states.
The values of the $0s$ and $0p$ orbits are the component of the harmonic oscillator basis wave functions in the $sp$ shells of the TOSM configuration.
The values of the $1s$ and $1p$ orbits are the other components except for the $0s$ and $0p$ ones, namely particle states in TOSM. They include the high-momentum components. 
From the table, the numbers of the $0s$ components are almost constant and the $0p_{1/2}$ and $0p_{3/2}$ components depend on the states.

\section{Summary}\label{sec:summary}

The nucleon-nucleon ($NN$) bare interaction has a strong tensor force at long and intermediate distances and a strong central repulsion at short distances.
We have treated the above characteristics of the $NN$ tensor force in terms of the tensor-optimized shell model (TOSM), in which 2p2h states are fully optimized to describe the deuteron-like tensor correlation. 
The short-range repulsion in the $NN$ interaction is treated using the central correlation of UCOM. 
In this study, we have investigated the structures of self-conjugate $4N$ nuclei, $^4$He and $^8$Be.
We mainly focus on the different structures appearing in the ground and the excited states of $^8$Be for the $T$=0 and $T$=1 states.
Experimentally, the $^8$Be nucleus shows two kinds of interesting aspects of $\alpha$ clustering in the ground band states and the highly excited states
in which the $\alpha$ decay process is not necessarily favored.

We have applied TOSM to $^8$Be to investigate these totally different structures of $^8$Be in relation to the tensor force.
For this purpose, we newly define the effective interaction for TOSM based on the AV8$^\prime$ interaction, 
in which the strengths of the tensor and $LS$ forces are increased in order to simulate the few-body calculation of $^4$He as a reference nucleus.
Hence this effective interaction retains the characteristics of the bare $NN$ interaction.
This prescription nicely recovers the missing strengths of the tensor and $LS$ contributions in $^4$He,
which mainly come from the coupling between the short-range UCOM transformation and the short-range parts of the tensor and $LS$ correlations.

It is found that TOSM reproduces fairly well the excitation energy spectrum of $^8$Be, except for the energy spacing between the ground band states and the highly excited states.
The resulting small energy spacing is related to the lack of the $\alpha$ clustering component in the ground band states in TOSM.
The three states belong to a rotational band and their Hamiltonian components show larger kinetic energy, central, and tensor contributions than other states.
In particular, the ground state possesses almost twice the $^4$He values for the kinetic energies and central contributions, which could be the signature of two-$\alpha$ clustering in $^8$Be. 
The tensor contributions do not reach the twice of the $^4$He value, where $^4$He contains the 2p2h states induced by the tensor force in TOSM.
This fact suggests the necessity of the higher configurations in $^8$Be such as 4p4h states in TOSM, to express the two-$\alpha$ clustering component involving the strong tensor correlation in each $\alpha$ cluster.

For highly excited states, we normalize the energy spectrum to the $2^+$($T$=1) state,
because this state is the isobaric analog state of the $^8$Li ground state and TOSM was able to  describe the structures of Li isotopes in the previous study.
This normalization of the $^8$Be spectrum makes it easier to understand the energy locations of the ground band states and the highly excited states obtained in TOSM in comparison with the experiment.
It is found that TOSM provides a good level order to reproduce the experiments for both the $T$=0 and $T$=1 states.
This result indicates that the highly excited states of $^8$Be can be regarded as shell-like states that TOSM can treat these states very well.
On the other hand, when we employ the effective Minnesota interaction without the tensor force, the results show a different energy level order.
Hence, the state dependence of the tensor force is necessary for $^8$Be, which is correctly treated in TOSM using the bare interaction.
It is also shown that the $T$=0 states in $^8$Be generally possess stronger tensor contributions than the $T$=1 states.
This is naturally understood from the attractive properties of the $T$=0 channel of the tensor force.

To understand the roles of the non-central forces explicitly, 
we have examined the dependences of the tensor and $LS$ matrix elements on the $^8$Be structures.
The tensor force mainly affects the degeneracy of the energies of the $T$=0 and $T$=1 states in the highly excited states of $^8$Be. 
We also discuss the effect of the $LS$ force, which contributes to determining the level spacing in the $^8$Be spectrum.

$\alpha$ clustering is an important aspect in $^8$Be and we estimate the correlation energy of two-$\alpha$ clustering, 
which is not fully included in the present TOSM using the bare $NN$ interaction.
We have used Brink's $\alpha$ cluster model and obtained 4 MeV as the contribution for sufficient $\alpha$ clustering.
This energy can recover the small energy spacing between the ground band states and the highly excited states in TOSM.

From the analysis with TOSM, two kinds of structures of $^8$Be in the three ground band states and the highly excited states 
are clarified in relation to the tensor force and the $\alpha$ clustering.
In particular, TOSM successfully describes the shell-like states in the highly excited states. 
For $\alpha$ clustering states in the three ground band states, TOSM is not sufficient to describe them and 
we have calculated the correlation energy from the two-$\alpha$ clustering component. 
We have discussed the necessary components for the $\alpha$ clustering in TOSM.
In $^8$Be, the $\alpha$ cluster and shell-like states are located at lower and higher excitation energies, respectively.
It would be interesting to extend the present discussion to other $4N$ nuclei such as $^{12}$C and $^{16}$O, which have the similar aspects of the shell and $\alpha$ cluster structures.
In these nuclei, the ground states show shell-like states and the excited states such as the Hoyle state in $^{12}$C, exhibit clustering structures,
which is the opposite situation to the $^8$Be case in the excitation energy. 
A unified description of those different structures in each nuclei is desirable and,
for this purpose, TOSM should be extended to treat the asymptotic condition which expresses the spatial localization of the $\alpha$ clusters.

\section*{Acknowledgment}
We thank Professor Hisashi Horiuchi for fruitful discussions and continual encouragement.
This work was supported by Grant-in-Aids for Scientific Research from the Japan Society for the Promotion of Science (C) 24740175 and (S) 23224004.
Numerical calculations were performed on a computer system at RCNP, Osaka University.

\def\JL#1#2#3#4{ {{\rm #1}} \textbf{#2}, #4 (#3)}  
\nc{\PR}[3]     {\JL{Phys. Rev.}{#1}{#2}{#3}}
\nc{\PRC}[3]    {\JL{Phys. Rev.~C}{#1}{#2}{#3}}
\nc{\PRA}[3]    {\JL{Phys. Rev.~A}{#1}{#2}{#3}}
\nc{\PRL}[3]    {\JL{Phys. Rev. Lett.}{#1}{#2}{#3}}
\nc{\NP}[3]     {\JL{Nucl. Phys.}{#1}{#2}{#3}}
\nc{\NPA}[3]    {\JL{Nucl. Phys.}{A#1}{#2}{#3}}
\nc{\PL}[3]     {\JL{Phys. Lett.}{#1}{#2}{#3}}
\nc{\PLB}[3]    {\JL{Phys. Lett.~B}{#1}{#2}{#3}}
\nc{\PTP}[3]    {\JL{Prog. Theor. Phys.}{#1}{#2}{#3}}
\nc{\PTPS}[3]   {\JL{Prog. Theor. Phys. Suppl.}{#1}{#2}{#3}}
\nc{\PRep}[3]   {\JL{Phys. Rep.}{#1}{#2}{#3}}
\nc{\AP}[3]     {\JL{Ann. Phys.}{#1}{#2}{#3}}
\nc{\JP}[3]     {\JL{J. of Phys.}{#1}{#2}{#3}}
\nc{\andvol}[3] {{\it ibid.}\JL{}{#1}{#2}{#3}}
\nc{\PPNP}[3]   {\JL{Prog. Part. Nucl. Phys.}{#1}{#2}{#3}}
\nc{\FBS}[3]   {\JL{Few Body Syst.}{#1}{#2}{#3}}

\end{document}